\documentclass[aps,showpacs,preprint,superscriptaddress]{revtex4}
\usepackage{graphicx}
\usepackage{amsmath}
\usepackage{bm}

\begin{document}

\title{Enhanced correlation of electron-positron pair in the two and three dimensions}
\author{Suo Tang}
\author{Bai-Song Xie\footnote{Corresponding author. Email address: bsxie@bnu.edu.cn}}
\affiliation{Key Laboratory of Beam Technology and Materials Modification of the Ministry of Education, College of Nuclear Science and Technology, Beijing Normal University, Beijing 100875, China}
\author{Hong-Yu Wang}
\affiliation{Department of Physics, Anshan Normal University, Anshan 114005, China}
\author{Jie Liu}
\author{Li-Bin Fu}
\affiliation{Institute of Applied Physics and Computational Mathematics, P. O. Box 8009, Beijing 100088, China}
\author{M. Y. Yu}
\affiliation{Institute for Fusion Theory and Simulation, Department of Physics,
Zhejiang University, Hangzhou 310027, China}
\affiliation{Institut f\"ur Theoretische Physik I, Ruhr-Universi\"at Bochum, D-44780 Bochum, Germany}

\begin{abstract}
Early-time electron-positron correlation in vacuum pair-production in an external field is investigated. The entangled electron and positron wave functions are obtained analytically in the configuration and momentum spaces. It is shown that, relative to that of the one-dimensional theory, two- and three-dimensional calculations yield enhanced spatial correlation and broadened momentum spectrum. In fact, at early times the electron and positron almost coincide spatially. The correlation also depends on the direction of the applied field. For the spatial correlation, the transverse correlation is stronger than the longitudinal one. \end{abstract}


\maketitle

Pair creation in vacuum can be considered as ``vacuum breakdown" by a supercritical field \cite{W.Greiner.B.muller}. Since Sauter \cite{F.Sauter} and Schwinger \cite{J.Schwinger} obtained the pair creation rate in a static field, many theoretical and experimental studies on this topic have been performed \cite{PRL17,PRL193902,PRL043004,PRL1626,PRL105001,PRL105003}. It has also been shown that the pair creation rate can be improved by tailoring the applied field \cite{PRL140402,PRA053402,PRL165006,PRA012106}. However, the pair birth process itself, such as the space and momentum correlations between the just-created electrons and positrons, is less studied \cite{PRL043004,PRL040406,J.mod.Opt}.

In this Letter, using computational quantum field methods \cite{PRA604} we investigate the electron-positron joint probability distributions in the configuration and momentum spaces. As in Ref.\ \onlinecite{PRL043004}, we are interested in the very early stage, namely at times $t\ll 1/c^2$ after the applied potential is turned on. 
The one-dimensional (1D) spatial density distribution \cite{PRL043004} is generalized to higher dimensions. The correlated spatial and momentum distributions parallel and perpendicular (referred to as the longitudinal and transverse correlations, respectively) to the external field are obtained.

In our model, we use the Sauter potential \cite{F.Sauter} $V(\bm{r})=V_0[1+\tanh(x/W)]/2$, where $W$ is the spatial extent of the corresponding electric field. The potential is abruptly turned on at $t=0$. 
The evolution of the field operator $\hat{\Psi}(\bm{r},t)$ is given by the Dirac equation \cite{S.S.Schweber} $i\partial_{t}\hat{\Psi}(\bm{r},t)=(c\bm{\alpha}\hat{\bm{P}}+\beta c^2+V)\hat{\Psi}(\bm{r},t)$, where $\bm{\alpha}$ and $\beta$ are the Dirac matrices, $c$ is the vacuum light speed, $\hat{\bm{P}}$ is the momentum operator, and $V$ is the external potential. The field operator can be written in terms of the electron creation and annihilation operators as
\begin{align}
\hat{\Psi}(\bm{r},t)&=\sum_{\bm{p}}\hat{b}_{\bm{p}}(t)W_{\bm{p}}(\bm{r})+\sum_{\bm{n}}\hat{d}^{\dag}_{\bm{n}}(t)W_{\bm{n}}(\bm{r})  \notag \\
&=\sum_{\bm{p}}\hat{b}_{\bm{p}}W_{\bm{p}}(\bm{r},t)+\sum_{\bm{n}}\hat{d}_{\bm{n}}^{\dag}W_{\bm{n}}(\bm{r},t),
\label{eq1}
\end{align}
where $W_{\bm{p}(\bm{n})}(\bm{r})$ is the energy eigenfunction of the field-free Dirac equation, and $W_{\bm{p}(\bm{n})}(\bm{r},t)=\langle\bm{r}|U(t)|\bm{p}(\bm{n})\rangle$ is the solution of the time-dependent Dirac equation with the time evolving operator $U(t)=\exp[-i(c\bm{\alpha}\hat{\bm{P}}+\beta c^2+V)t]$. The electron-positron wave function is given by the positive-frequency parts of the field operator and its charge-conjugated field operator $\phi(\bm{r}_1,\bm{r}_2,t)=\langle 0|\hat{\Psi}^{(+)}(\bm{r}_1,t)\bigotimes\hat{\Psi}^{(+)}_{c}(\bm{r}_2,t)|0\rangle$, here $\bm{r}_1$ and $\bm{r}_2$ are the spatial coordinates of the created electron and positron, respectively. Unless otherwise stated, in this paper atomic units ($e=m_e=\hbar=1$) are used. Based on the eigenstates of the free Dirac operator we can express the pair wave function as \cite{PRL043004}
\begin{equation}
\phi(\bm{r}_1,\bm{r}_2,t)=\sum_{\bm{n}}\sum_{\bm{p}}A_{\bm{pn}}(t)W_{\bm{p}}(\bm{r}_1)\bigotimes CW^{*}_{\bm{n}}(\bm{r}_2),
\label{eq2}
\end{equation}
where the matrix $C$ is the charge-conjugated operator and $A_{\bm{pn}}(t)=\sum_{\bm{P}}\langle\bm{p}|U(t)|\bm{P}\rangle\langle\bm{n}|U(t)|\bm{P}\rangle^{*}$ is the expansion coefficient.

For $t\ll 1/c^2$, up to $O(t^2)$ accuracy one obtains $A_{\bm{pn}}=i\langle\bm{p}\mid V\mid\bm{n}\rangle t=iV_{\bm{pn}}t$. Thus, the wave function $\phi(\bm{r}_1,\bm{r}_2,t)=\phi_0(\bm{r}_1,\bm{r}_2)t$ grows linearly with time. One can then explore the pair creation process in this early-time regime by following the reduced time-independent wave function $\phi_0(\bm{r}_1,\bm{r}_2)=i\sum_{\bm{n}}\sum_{\bm{p}}V_{\bm{pn}}W_{\bm{p}}(\bm{r}_1)\otimes CW^{*}_{\bm{n}}(\bm{r}_2)$. To investigate the spatial entanglement between the electron and positron, we consider the spatial joint probability distribution $\rho(\bm{r}_1,\bm{r}_2)=\mid \phi_0(\bm{r}_1,\bm{r}_2)\mid^2$ for the electron at $\bm{r}_1$ and the positron at $\bm{r}_2$. The probability of finding the pair is given by \cite{laser physics} $P(t)=t^2\int d\bm{r_1}d\bm{r_2}\rho(\bm{r}_1,\bm{r}_2)=t^2\int d\bm{p}d\bm{n}|V_{\bm{pn}}|^2$. The corresponding spectrum of the spatial joint probability distribution, i.e., the momentum joint probability distribution, is $\rho(\bm{p},\bm{n})=|V_{\bm{pn}}|^2$, which measures the electron with momentum $\bm{p}$ and the positron with momentum $\bm{n}$.

According to the Dirac theory, the wave functions of the free eigenstates are
\begin{equation}
W_{\bm{p}}(\bm{r})=\frac1{(2\pi)^{3/2}}\sqrt{\frac{c^2}{E_{\bm{p}}}}\mu_{\bm{p}}\exp(i\bm{pr}),
\label{eq3}
\end{equation}
with the eigenvalue $E_{\bm{p}}=\sqrt{c^2p_x^2+c^2p_y^2+c^2p_z^2+c^{4}}$ and
\begin{equation}
W_{\bm{n}}(\bm{r})=\frac1{(2\pi)^{3/2}}\sqrt{\frac{c^2}{E_{\bm{n}}}}\nu_{\bm{n}}\exp(-i\bm{nr}),
\label{eq4}
\end{equation}
with the eigenvalue $-E_{\bm{n}}=-\sqrt{c^2n_x^2+c^2n_y^2+c^2n_z^2+c^{4}}$. Here, $\mu_{\bm{p}}$ and $\nu_{\bm{n}}$ are the Dirac 4D spinors of the electron and positron, respectively. Because the external potential $V$ depends only on $x$, the simplified expansion coefficient $V_{\bm{pn}}=\langle \bm{p}\mid V(x)\mid\bm{n}\rangle$ should contain the factors $\delta(p_y+n_y)$ and $\delta(p_z+n_z)$, which serve to ensure momentum conservation in the transverse directions. For simplicity, for the 2D case we can choose the spinors as \cite{Wave Equations} $\mu_{\bm{p}}=\sqrt{(E_{\bm{p}}+c^2)/2c^2}(1,0,0,cp_{+}/(E_{\bm{p}}+c^2))$ and $\nu_{\bm{n}}=\sqrt{(E_{\bm{n}}+c^2)/2c^2}(cn_{-}/(E_{\bm{n}}+c^2),0,0,1)$, where $p_{+}=p_x+ip_y$ and $n_{-}=n_x-in_y$. There is no need to take all the spinors into account. Because the nonzero $8$ components of $4\times 4=16$ entangled wave function can be divided into two sets of $4$ components form and they are conjugate with each other in the 2D system. So without lose validity we need only one set of $4$ components, e.g. $\phi_0^i(\xi_x,\xi_y)$, where $i=1,2,3,4$. In the limit of $W=\infty$, the potential approximates to $V(x,y)=V_0(1+x/W)/2$. The 2D reduced wave function $\phi_0(\xi_x,\xi_y)$, (here and throughout the paper we omit $i$, which does not cause any confusion) is then
\begin{align}
\phi_0(\xi_x,\xi_y)&=\frac{V_0c}{32W\pi^2}\int dP_x dP_y \frac{E_{P_xP_y}+1+P_y^2+iP_xP_y}{E_{P_xP_y}^{3}}\notag \\
& \times
\left(\begin{array}{ccc}
1\\
-\frac{P_x+iP_y}{1+E_{P_xP_y}}\\
\frac{P_x+iP_y}{1+E_{P_xP_y}}\\
-\left(\frac{P_x+iP_y}{1+E_{P_xP_y}}\right)^2
\end{array}\right)
\exp(iP_x\xi_x)\exp(iP_y\xi_y),
\label{eq5}
\end{align}
where $E_{P_xP_y}=\sqrt{1+P_x^2+P_y^2}$ and $\xi_x=c(x_1-x_2)$, $\xi_y=c(y_1-y_2)$, and $P_x=(p_x-n_x)/2c$ and $P_y=p_y/c$ are the relative momenta in the $x$ and $y$ directions measured by $c$. Clearly, except for a constant factor the momentum-space wave function $\phi_0(P_x,P_y)$ can be regarded as the Fourier transformation of $\phi_0(\xi_x,\xi_y)$, or
\begin{align}
\phi_0(P_x,P_y)&=\frac{E_{P_xP_y}+1+P_y^2+iP_xP_y}{E_{P_xP_y}^{3}}
\left(\begin{array}{ccc}
1\\
-\frac{P_x+iP_y}{1+E_{P_xP_y}}\\
\frac{P_x+iP_y}{1+E_{P_xP_y}}\\
-\left(\frac{P_x+iP_y}{1+E_{P_xP_y}}\right)^2
\end{array}\right),
\label{eq6}
\end{align}
so that the momentum spectrum is
\begin{equation}
\rho(P_x,P_y)=|\phi_0(P_x,P_y)|^2=4\frac{1+P_y^2}{(1+P_x^2+P_y^2)^2}.
\label{eq7}
\end{equation}

\begin{figure}
  \includegraphics[width=6cm]{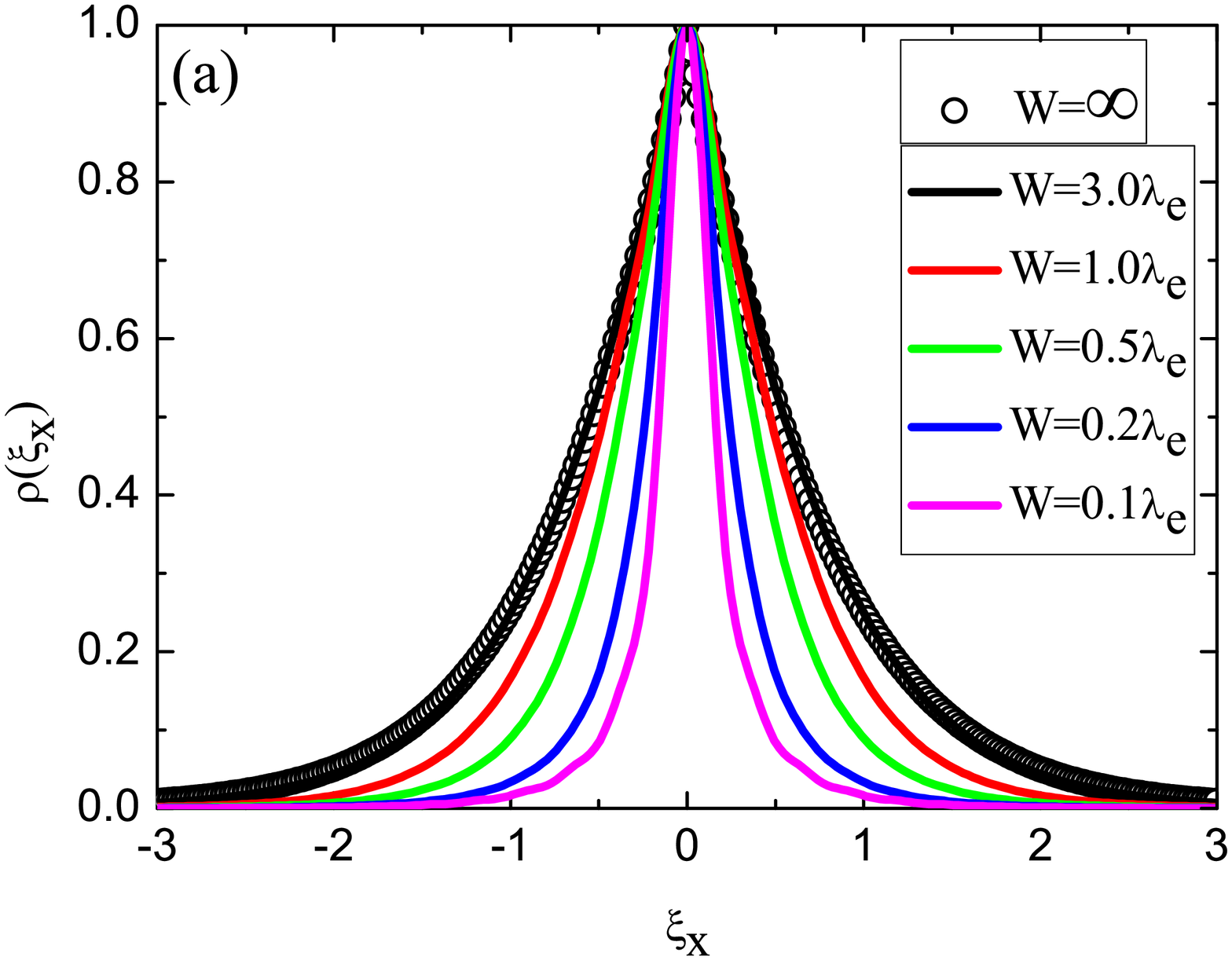}
  \includegraphics[width=6cm]{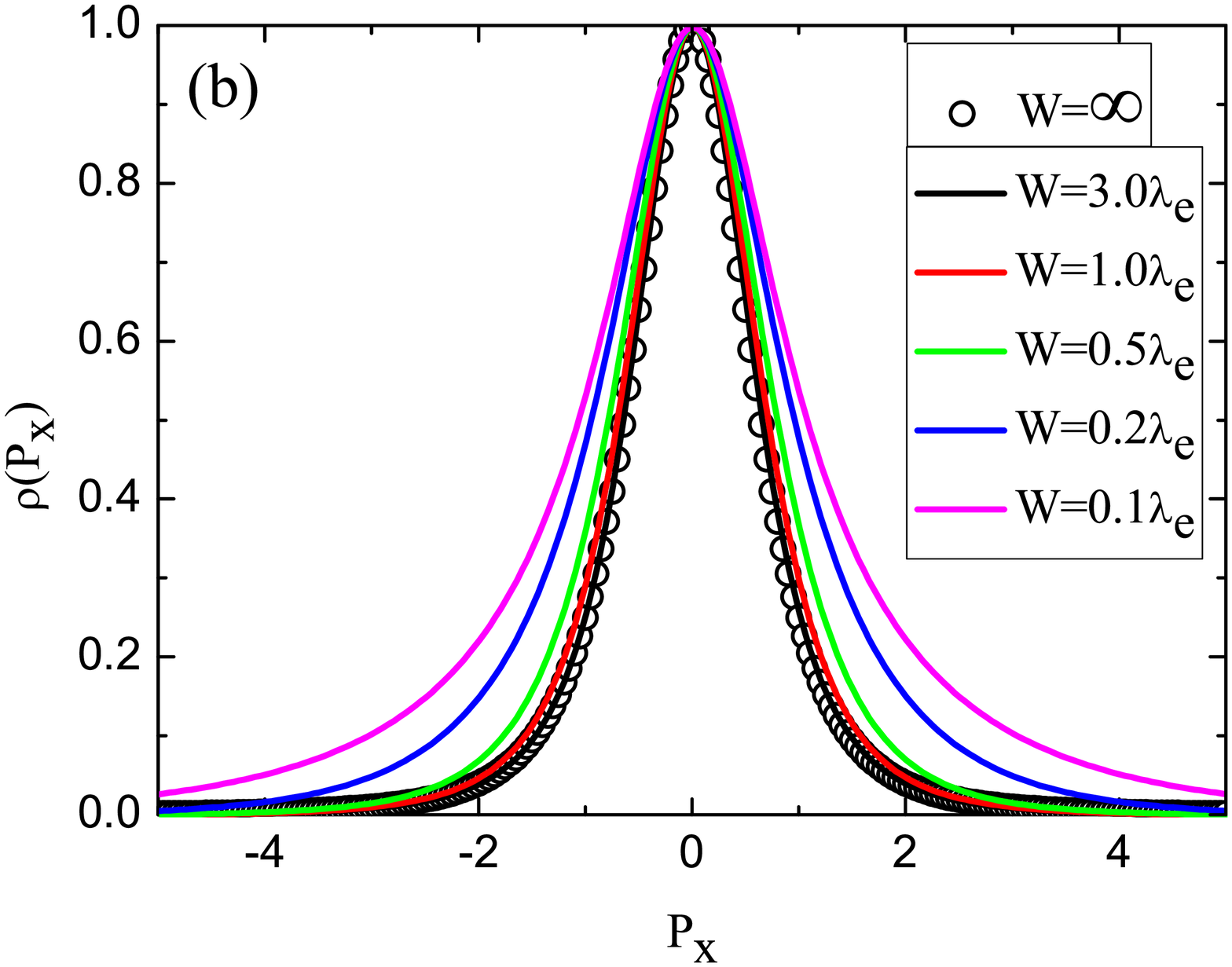}
  \caption{(Color online) (a) The spatial density distribution $\rho(\xi_x)$ for the 1D case. (b) The corresponding momentum spectrum $\rho(P_x)$. The circles represent the results for $W=\infty$. The data are scaled to match at $\xi_x=0$ and $P_x=0$ for better eye view.}
  \label{fig.1.}
\end{figure}

To see in more detail the relation between the joint distributions in configuration and momentum spaces, we first consider a 1D system \cite{PRL043004} by setting $p_y=n_y=0$. Without loss of generality, we can assume that the positron is always detected at $x=0$. In Fig.\ \ref{fig.1.}, we show the joint distribution in the presence of fields of different widths. As one can see, if the spatial extent of the field is wider than $3\lambda_e $, where $\lambda_e $ is the electron Compton wavelength, the joint distributions agree with the analytical result for an infinite-width field, shown as circles in Fig.\ \ref{fig.1.}. As the field width becomes narrower, the average distance between the particles shrinks until it vanishes \cite{PRL043004}. As expected, the corresponding momentum spectrum broadens until the width becomes constant.

\begin{figure}
  \includegraphics[width=6cm]{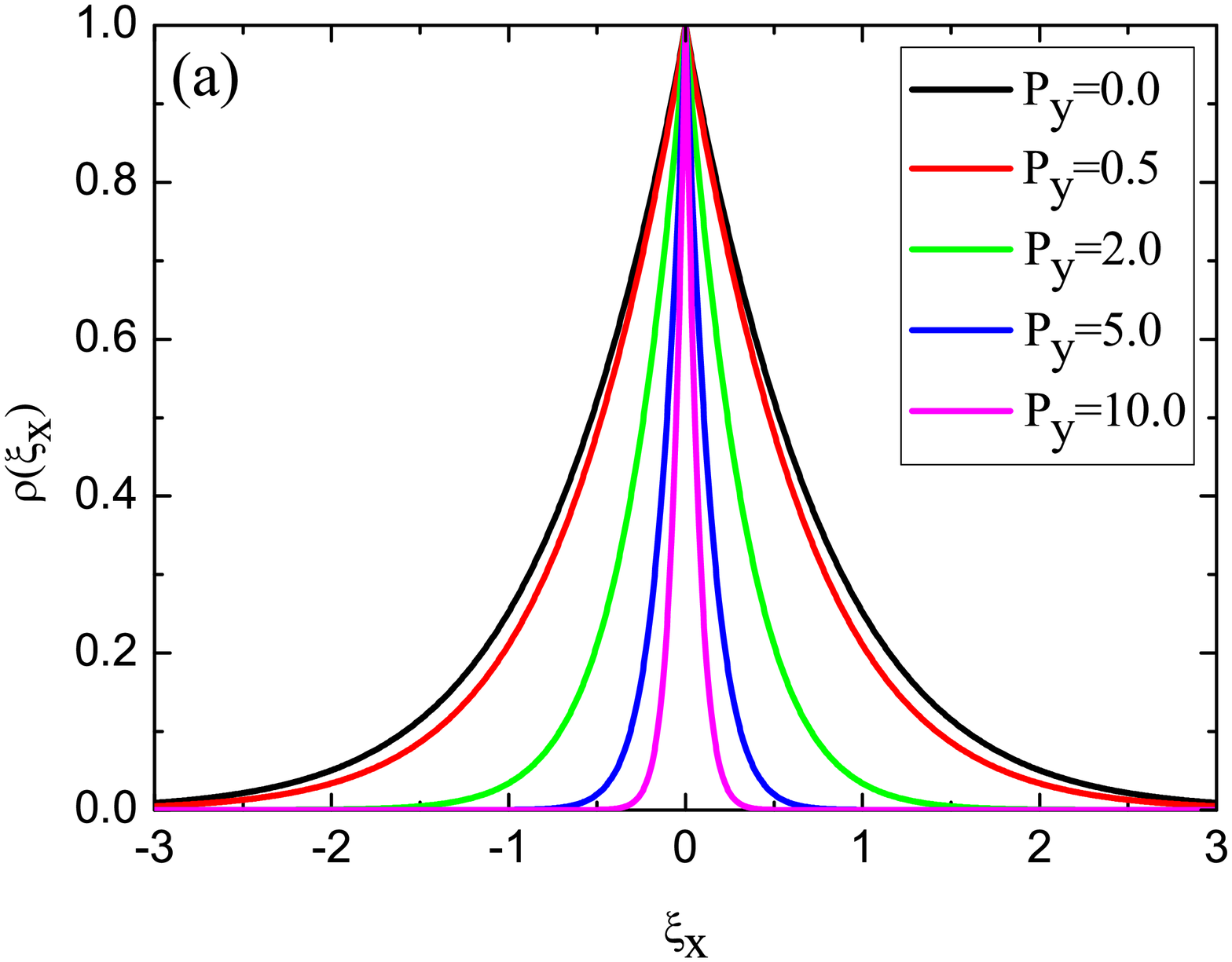}
  \includegraphics[width=6cm]{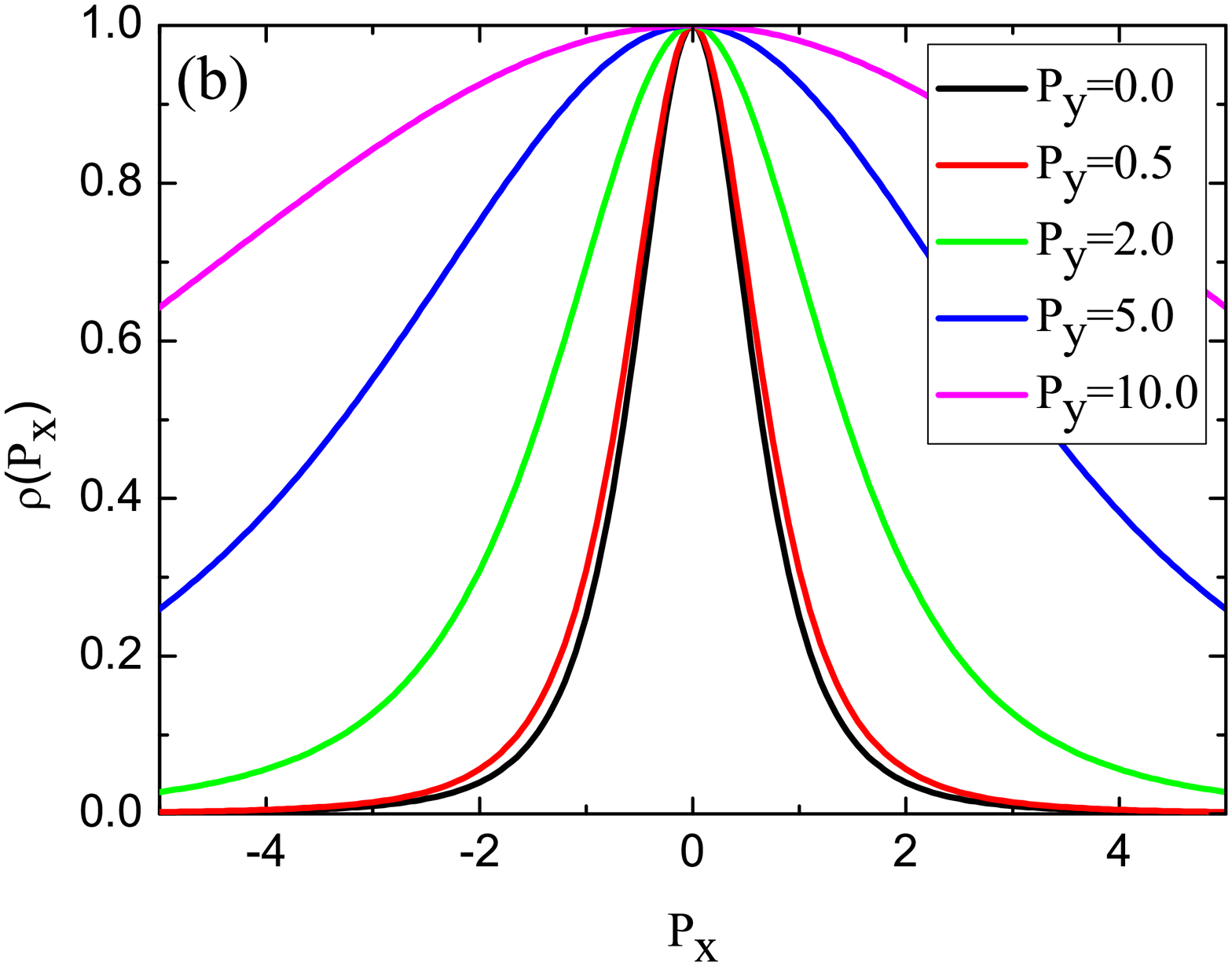}
  \caption{(Color online) The electron-positron spatial density distribution $\rho(\xi_x)$ in (a) and the corresponding momentum spectrum $\rho(P_x)$ in (b) in the quasi-2D case for different given $P_y$. The analytical wave function Eqs.\ (\ref{eq5}) and (\ref{eq6}) are used. The data are scaled to match at $\xi_x=0$ and $P_x=0$.}
  \label{fig.2.}
\end{figure}

Fig.\ \ref{fig.2.} shows the electron joint distribution obtained from the analytical wave functions Eq.s\ (\ref{eq5}) and (\ref{eq6}) for a quasi-2D system \cite{PRA013422} with finite $y$ momenta. We see that as $p_y$ increases the spatial density distribution in the $x$ direction becomes narrower, accompanied by broadening of the corresponding momentum spectrum, as expected since the spatial distribution width $\Delta x$ and the momentum spectrum width $\Delta P_{x}$ should satisfy the uncertainly principle, $\Delta x\Delta P_{x}\sim 1$. It can also be verified that in the 1D limit ($p_y=0$), both the spatial and momentum distributions agree with that shown in Fig.\ \ref{fig.1.}.

Now we consider a full 2D system. For a field of infinite width we can express the spatial joint density as $\rho(\xi_x,\xi_y)=|\phi_0(\xi_x,\xi_y)|^2$. The longitudinal and transverse joint densities in different directions are given by $\rho(\xi_x)=\int d\xi_y \rho(\xi_x,\xi_y)$ and $\rho(\xi_y)=\int d\xi_x \rho(\xi_x,\xi_y)$. Since $\rho(\xi_x=0,\xi_y=0)$, $\rho(\xi_x=0)$, and $\rho(\xi_y=0)$ are all divergent, i.e., $(0,0)$ is a singular point of the density $\rho$, we cannot scale the spatial density distributions so that they match at $\xi_x=0$ or/and $\xi_y=0$, as done above. On the other hand, Eq.\ (\ref{eq7}) shows that in the dual space we can still obtain the momentum spectrum, which qualitatively reflects the properties of the spatial density. The longitudinal and transverse components of the momentum spectrum in different directions are
\begin{align}
\rho(P_x)&=\int dP_y\rho(P_x,P_y)=\frac{2\pi}{(1+P_x^2)^{1/2}}+\frac{2\pi}{(1+P_x^2)^{3/2}},
\notag \\
\rho(P_y)&=\int dp_x\rho(P_x,P_y)=\frac{2\pi}{(1+P_y^2)^{1/2}}.
\label{eq8}
\end{align}

\begin{figure}
  \includegraphics[width=12cm]{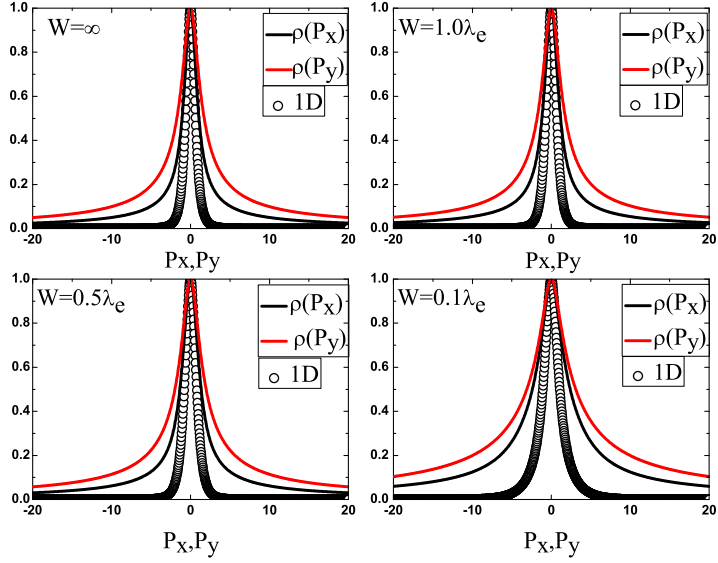}
  \caption{(Color online) The momentum spectrum $\rho(P_x)$ and $\rho(P_y)$ in the 2D case for different $W$. The circles are the spectrums in the 1D case. The data are scaled to match at $(P_x,P_y)=0$.}
  \label{fig.3.}
\end{figure}

\begin{figure}
  \includegraphics[width=8cm]{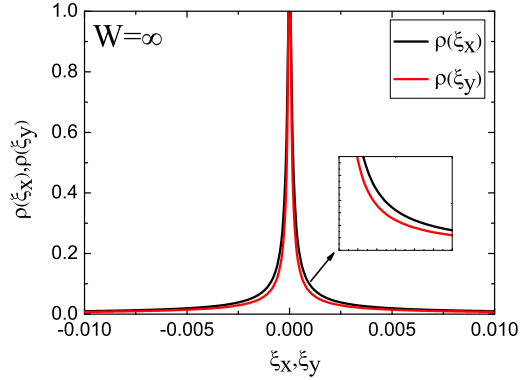}
  \caption{(Color online) The density distributions $\rho(\xi_x)$ and $\rho(\xi_y)$ in the 2D case for different $W=\infty$. The data are not scaled to match at $\xi_x, \xi_y=0$ because the divergence at this point. Instead they are scaled to match at $\xi_{x},\xi_{y}=0.00005$.}
  \label{fig.4.}
\end{figure}

In Fig.\ \ref{fig.3.} we show the momentum spectrum in the 2D system for different field widths. For comparison, the corresponding distributions in a 1D system are also shown (circles).  Eq.\ (\ref{eq8}) is used for the field width $W=\infty$. The distributions corresponding to the other field widths are obtained numerically. One sees that the momentum spectra in the 2D system are much wider than that in the 1D system. We can thus expect that the density distribution in 2D is much narrower than that in 1D. That is, in the 2D system the electron-positron momentum correlation is weakened, but the spatial correlation is enhanced. Moreover, we note that the transverse spectrum is always wider than the longitudinal one, so that one can expect that $\rho(\xi_y)$ is always narrower than $\rho(\xi_x)$, or, the electron-positron spatial correlation perpendicular to the external field is stronger. In order to verify these conclusions, the longitudinal and transverse density distributions of the 2D system are obtained directly by numerically solving Eq.\ (\ref{eq5}) together with the definitions of $\rho(\xi_x)$ and $\rho(\xi_y)$. The results are shown in Fig.\ \ref{fig.4.}. Clearly, both of $\rho(\xi_x=0)$ and $\rho(\xi_y=0)$ are divergent. Moreover, we can clearly see that their widths are narrower than that in the 1D case and that the width of $\rho(\xi_y)$ is narrower than that of $\rho(\xi_x)$.


Similarly, for the 3D system we can obtain analytically for the field of width $W=\infty$ the joint momentum spectrum
\begin{equation}
\rho(\bm{P})=8\frac{1+P_y^2+P_z^2}{(1+P_x^2+P_y^2+P_z^2)^2},
\label{eq9}
\end{equation}
where $P_z=p_z/c$ and $\bm{P}=(P_x,P_y,P_z)$ and all the four spinors are taken into account. We see that $\rho(P_x)/\rho(P_x=0)\to 1$, so that the width of the momentum spectrum in the parallel direction approaches infinity, $\triangle P_{x}\rightarrow  \infty$, and the width of the spatial distribution approaches zero, $\triangle \xi_{x}\rightarrow  0$. The same situation occurs for the perpendicular directions, namely, $\rho(P_y)/\rho(P_y=0)\to 1$, $\rho(P_z)/\rho(P_z=0)\to 1$ and $\triangle \xi_{y}\rightarrow  0$, $\triangle \xi_{z}\rightarrow  0$. Accordingly, in the 3D system the electron and positron are pair-created at practically the same location. This result differs strongly from that of the 1D case, where the average spatial extent of pair creation is of order $\lambda_e$.

We can see why the spatial correlation is enhanced in the high dimensional system. Because more information of a particle can be gained by a position measurement of the accompanying particle \cite{J.mod.Opt}. The enhancement comes from the coupling influence of the momenta in the different directions which can be illuminated from Eqs.\ (\ref{eq5}) and (\ref{eq6}). For example in the 2D case the transverse momentum makes the longitudinal momentum spectrum broaden with the corresponding momentum distribution approaches zero slowly as momentum increases compared to the 1D case. In the other words the coupling of entangled wave function due to different dimensions leads to that the spatial correlation degree is increased greatly.

In summary, we have considered electron-positron correlations in the pair-creation process during the early stage. Our results suggest that the correlations in configuration and momentum spaces exhibit reciprocal duality which is consistent with the uncertainty principle, i.e. $\triangle P_{x} \triangle x \approx 1$. It is found that the correlations depend on the dimension of the computational space. Concretely, the particle pair are created at almost the same location in the 3D system while the average distance between the particles is finite in lower dimensional spaces for the infinite width field. We also find the transverse spatial correlation is stronger than the longitudinal one in the 2D system and in the quasi-2D system, the longitudinal spatial correlation increases with the transverse momentum.

Authors acknowledge helpful discussions with Professors Shigang Chen and Xinheng Guo. This work was supported by the National Natural Science Foundation of China (11175023 and 11247007) and partially by the Open Fund of National Laboratory of Science and Technology on Computational Physics, IAPCM, the Open Fund of the State Key Laboratory of High Field Laser Physics, SIOM, and the Ministry of Science and Technology of China (2011GB105000).

\end{document}